\documentstyle[11pt,newpasp,twoside,epsf]{article}
\def\edcomment#1{\iffalse\marginpar{\raggedright\sl#1\/}\else\relax\fi}
\reversemarginpar

\begin{document}
\title{The Thermal Sunyaev Zel'dovich effect : a powerful probe for missing
 baryons}
 \author{Pengjie Zhang}
\affil{Astronomy Department, Univ. of Toronto, Canada}
\author{Ue-Li Pen}
\affil{Canadian Institute for Theoretical Astrophysics}

\begin{abstract}
About $90 \%$ of baryons in the universe have thus far escaped direct
observation. This is known as the {\it missing baryon problem}. The
Sunyaev Zel'dovich effect (SZ effect) has the potential to directly measure the state of the
majority  of these missing baryons. The next generation CMB experiments
such as AMIBA will provide an unbiased sample of the intergalactic
medium through the SZ effect. The
existing and upcoming simulations and analytical studies provide a
quantitative  
understanding of the SZ effect. All these make the SZ 
effect a powerful baryon probe. We present an overview
of this probe from both phenomelogical and theoretical aspects.
\end{abstract}

\section{Introduction}
The universe is turning out to be quite complicated.  The favorite
cosmology postulates that a mysterious 
negative pressure field or a cosmological constant accounts for about $2/3$ of the total energy
budget. The remaining $2/3$  is believed to be dominantly cold dark matter. The
ordinary  baryonic matter only accounts for about $5\%$ of
the total matter in the universe. But the directly observed components
such as stars, interstellar medium, intracluster gas, etc., only
account for about $10\%$ of the  total baryon budget
(Persic \& Salucci 1992; Fukugita, Hogan \& Peebles 1998). This is the so called missing baryon 
problem. These missing baryons are  believed to be in 
the form of the intergalactic medium (IGM) and are elusive to detect. To
understand their state, such as density, temperature, peculiar
velocity, metalicity, etc., 
stands as a major challenge to both observation and theory, and is
crucial to understanding the  thermal  history of the universe, galaxy
formation, etc.

   The IGM has various observable tracers. (1) Neutral
hydrogen
absorbs the background light of Quasars and produces the Lyman-$\alpha$
forest. (2) The ionized electrons have thermal and peculiar motions and
are capable of scattering CMB photons and generate  secondary
CMB temperature fluctuation, which are known as the  thermal and
kinetic Sunyaev Zel'dovich effects (TSZ and KSZ effect),
respectively. Their precision measurements are  becoming 
routinely available with devoted CMB experiments such as  AMIBA. 
(3) The ionized electrons and protons interact
with each other and emit X-rays through thermal bremsstrahlung and
contribute to the soft X-ray ($0.5-2$ keV) background (XRB).

These effects have different dependences on the IGM state and probe
different phases. The Lyman-$\alpha$ forest probes 
the low density neutral IGM phase where the overdensity $\delta\equiv
\delta \rho/\rho -1 \sim 1$. Since our
universe is highly ionized,  the Lyman-$\alpha$ forest only directly
probes a tiny  
fraction of the missing baryons and radiative modelling is required to
extrapolate to the state of the bulk of the gas. The X-ray
emission is proportional to $\rho^2$ and primarily probes the densest
IGM regions.  X-ray flux is diluted by distance
and further by the expansion of
the universe. So the observed IGM XRB is mostly from the nearby
universe ($z \leq 0.5$) and it is difficult to look
very deep. X-ray absorption is another possibility (Perna \& Loeb 1998), but is
strongly temperature and metalicity dependent. In contrast to these
two effects, the SZ effect (both TSZ and KSZ) is a
universal and powerful probe of baryons. Firstly, it provides an unbiased
sampling of the missing baryons. All free electrons participate in
Thomson scattering and contribute to the SZ effect. The Thomson
optical depth from the epoch of reionization $z\sim 10$  to the
present is $\tau \sim 0.1$, which means that a few percent of  CMB photons have been scattered by electrons. Secondly, the Compton scattering does not depend
on redshift and is not affected by distance or the expansion of
the universe. So, the SZ effect can probe the  more distant universe. Thirdly,
the SZ effect has a weak dependence on
the gas density in contrast to the X-ray emission and thus probes a large
range of baryon. Fourthly, from a theoretical
viewpoint, the SZ effect is easier to model and simulate than the
other two effects. In 
observations, the next generation CMB experiments such as AMIBA are able to
measure the SZ effect with high  accuracy and cover
a large proportion of the sky. All these advantages suggest that the
SZ effect is 
a powerful probe for the missing baryons. Since the thermal SZ effect is
about an order of magnitude larger than the kinetic SZ effect, hereafter
we focus on the thermal SZ effect.

    The biggest shortcoming of the SZ effect is its lack of redshift
information.  Since scattering is independent of redshift, the SZ
effect on one hand allows a direct probe of the IGM to high redshift,
on the other hand makes it challenging to disentangle the
contributions arising from different redshifts.  We have proposed a
method to recover the space distribution and time evolution of the
IGM (Zhang \& Pen 2001) to take full advantage of the SZ effect.

    In this paper, we first describe the SZ effect, its observational
feasibility and our variational method to extract the 3-D information
of the IGM pressure power spectrum in section section 2. We outline our
theoretical model of the SZ effect in section section 3. We
conclude in section 4.

\section{Observation}
\label{sec:obs}
The Cosmic Microwave Background (CMB) scatters off all free electrons
through inverse Compton scattering, causing a change in the CMB
temperature and allows us to ``see'' the IGM through scattering, in
analogy to absorption.  This effect is known as the  Sunyaev
Zel'dovich effect. The temperature distortion caused by the TSZ effect
(Zeldovich \& Sunyaev 1969)
is:
\begin{equation}
\Theta(\hat{n})\equiv\frac{\Delta T_{CMB}(\hat{n})}{T_{CMB}}=-y(\hat{n})
\frac{xe^x}{e^{x}-1}\left[4-x/\tanh(x/2)\right] \equiv -2yS(x),
\end{equation}
where $\hat{n}$ is the direction on the
sky and $x\equiv h\nu/(kT_{CMB})$. Since the typical temperature of the IGM 
$\sim 10^7$ K is much higher than $T_{CMB}=2.73 $K, the photons always
gain energy and are 
kicked up to higher frequencies. Thus, the number density of photons on
the (less energetic) left side of the CMB peak  decreases and results
in an apparent cooling of the CMB temperature in the Rayleigh-Jeans regime. From photon number
conservation, the number density of photons on the right side increases and
results in a heating. With a multiple frequency SZ survey, this
frequency dependence enables us to  distinguish the SZ effect from the
primary and secondary CMB anisotropies (Cooray, Hu \& Tegmark 2000).  

For a fixed frequency, what matters is the Compton $y$  parameter, which
is defined as 
\begin{equation}
y(\hat{n})=\frac{\sigma_T}{m_e c^2} \int_0^{l(z_{cmb})} n_e kT_g dl=
\frac{\sigma_T}{m_e c^2} \int P_e(\hat{n}) dl .
\end{equation}
Here, $T_g$ and $n_e$ are the temperature and number density of free electrons,
respectively.  $P_e$ is the gas pressure. $dl$ is the proper distance
along the path of CMB protons. The typical value of the temperature
distortion is a few $\mu$K. The SZ effect has been detected
for targeted clusters where $\Delta T \sim m$K. The next step is to
carry out a random sky SZ survey  to measure an unbiased statistical sample. 
The next generation  CMB experiments such  as AMIBA can realize this
goal.  In such observations, we observe the sky map of the CMB
temperature fluctuation and thus the angular power 
spectrum, which is defined by $C_l\equiv\langle 
\sum_{m=-l}^l a_{lm}a_{lm}^*\rangle/(2l+1)$ (Here,
$\Theta(\hat{q})\equiv\frac{\delta T}{T}(\hat{q})=\sum^{l,m} 
a_{lm}Y_{lm}(\hat{q})$). 
In the small angle approximation, 
\begin{equation}
\label{eqn:cl}
C_l=4 (\frac{\sigma_T}{m_e
c^2})^2\int_0^{x(z_{cmb})}P_{SZ}(k,z)|_{k=l/x} a^2 x(z){\rm d}x(z), 
\end{equation}
where $x(z)$ is the comoving distance and $P_{SZ}(k,z)$ is the power
spectrum of the gas 
pressure. $P_{SZ}$ contains detailed  information about the  IGM, but
is smeared by redshifts. Because the SZ effect is mostly
contributed by nonlinear structures at $z\sim 0.5-2$ (Zhang \& Pen 2001),
it should have a  strong cross correlation with galaxies in that
redshift range. Taking advantage of the cross correlations with a
photometric redshift survey such as Sloan Digital Sky Survey, we are
able to statistically extract 
IGM 3-D correlations and  evolution. The basic idea is sketched in
fig. 1.

We show that the galaxy selection function to maximize the SZ-galaxy cross
correlation coefficient is actually the ratio of the 3-D
pressure-galaxy cross correlation power spectrum and the galaxy
auto correlation power spectrum. With further investigation, we can
extract the pressure power spectrum (Zhang \& Pen 2001).

\section{Theoretical estimation}
\label{sec:theory}
   With all these promising upcoming observational capabilities, we need to
understand the SZ effect theoretically. One would like to predict the
value of the $y$ parameter, the shape of the SZ power
spectrum, the angular scale where it becomes dominant over the primary
CMB, contributions to $y$ and the power
spectrum from different redshifts, the dependence  on  the
cosmological parameters and the gas parameters and any other information about
the  intervening IGM that we can infer.

   Two analytical models have been considered to  answer these
questions. The first is the halo model method (see
Komatsu \& Kitayama 1999 for a review). The contribution to the SZ  
effect by a single halo is described by the gas distribution in the
halo, for example, an isothermal distribution. The collective effect is
integrated  by the Press-Schechter formalism.  

   The second approach is to directly deal with the underlying density
field as we proposed (Zhang \& Pen 2001). The gas temperature depends
on  the gravitational potential ($T\propto \phi$) and is thus
determined by the underlying density field. So roughly
$P_e \propto n_e\phi$. Then, we can
estimate the evolution of the mean gas pressure, which is shown in
fig. 2. Furthermore, the pressure correlation function  $\langle
P_e(x) P_e(x+r)\rangle \propto \langle 
n_e(x) \phi(x) n_e(x+r)\phi(x+r) \rangle$. With the hierarchical model of the
density field (Fry 1984, 
Scoccimarro \& Frieman 1999), we can decompose this high order density
correlation 
into the products of the 2-point density correlations. The result is
shown in fig. 3. Our calculation shows that, for a typical $\Lambda$CDM model,
(1)  the mean temperature distortion 
$\Delta T \sim 6 \times  10^{-6}$ K. (2) The SZ power spectrum exceeds the 
primary CMB anisotropy at $l\sim 2000$ and peaks around $l\sim 3000$
($\theta \sim 10'$) as shown in fig. 4. The position of the peak is an
unambiguous signature of the the underlying gas-dark matter correlation. (3) $C_l$ has a strong dependence on  $\sigma_8$ 
($C_l \propto \sigma_8^{6--9}$). (4) The dominant contribution to
$C_l$ depends on the angular scale,  varying from $z\sim 0.4$  at
$l\sim1000$ to $z\sim 1.5$ at $l\sim 10000$. Smaller angular scales probe the
more distant universe (fig. 5).

Non-gravitational heating  is a key ingredient needed to understand
the IGM (Pen 1999). We have estimated its effect and found that it will 
leave an unambiguous signature in the 3-D power spectrum
(Zhang \& Pen 2001). If non-gravitational heating is proportional to the
gravitational heating, the only signature is an enhancement  of the
amplitude of the power spectrum. A spatially homogeneous energy
injection  will produce an
uprising tail in the power spectrum towards smaller scales
(fig. 6). The non-gravitational
heating parameters including epoch and energy can be extracted from a
precise measurement of the 
SZ map in conjunction with a galaxy photometric redshift survey. 

\section{Conclusion}
\label{sec:conclusion}
We summarize the benefits of using the SZ effect as the probe for
the missing baryons:
\begin{itemize}
\item The SZ effect is based on the Thomson scattering, which is sensitive to the majority of baryons: the ionized
IGM. The interaction between photons and electrons (about
$10\%$ CMB photons are scattered) guarantees the SZ effect to be an
unbiased sample of the missing baryons.

\item Compton scattering is completely understood physics, so the SZ
effect can be cleanly computed from first principles. 

\item Compton scattering does not depend on redshift and thus the
SZ effect can readily probe the IGM at  high redshift.
\item The upcoming experiments such as AMIBA will provide a wealth of
information on the  SZ effect. The dependence of 
the SZ effect on frequency  allows the separation from
primary and other secondary CMB anisotropies.
\item With the aid of galaxy photometric redshift surveys and our
variational method, we can obtain the full time and space resolved IGM
state.
\item We can reasonably estimate the SZ effect in theory both
analytically and in simulations from first principles.
\item The SZ effect can help us resolve some key questions on IGM such
as the role of non-gravitational heating, which is of key importance
to galaxy formation.
\end{itemize}

The SZ effect encodes key cosmological information, is observationally
feasible and theoretically tractable. It  will allow us to uncover the
state of the missing baryons.

\begin{figure}
\label{fig:variation}
\plotone{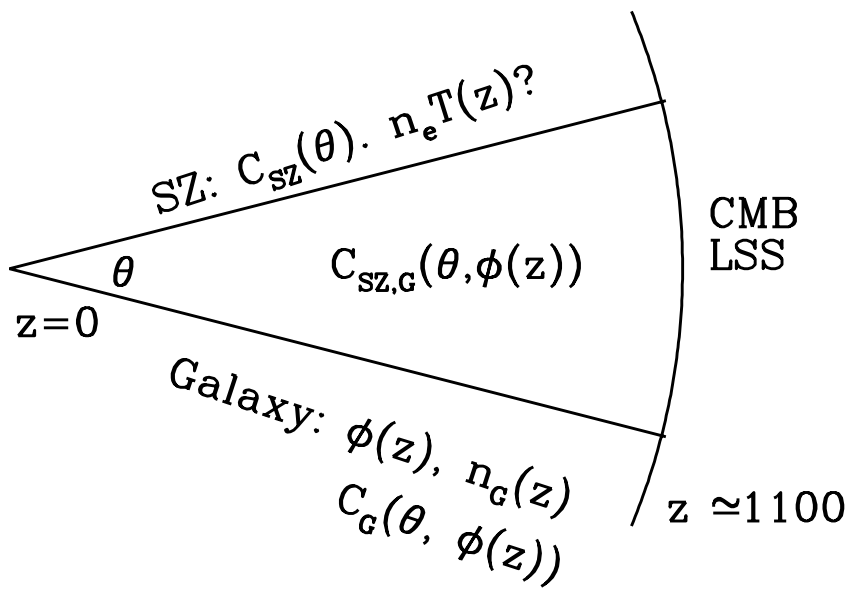}
\caption{Variational method to extract redshift information. The
observables are the SZ angular correlation $C_{SZ}$, galaxy angular
correlation function $C_G$, their cross correlation $C_{SZ,G}$ and the
3-D galaxy overdensity 
correlation. We want to extract the redshift distribution of
the IGM pressure $P_e=n_e k_B T$ and its correlations. The freedom in this
data set is the galaxy selection function $\phi(z)$. We vary this
function and maximize $\frac{C_{SZ,G}(\phi)}{\sqrt{C_{SZ}C_G(\phi)}}$. The
optimal $\phi_M(z)$ obtained in this way explicitly tells us the
underlying 3D pressure distribution}.
\end{figure}

\begin{figure}
\label{fig:meanP}
\plotone{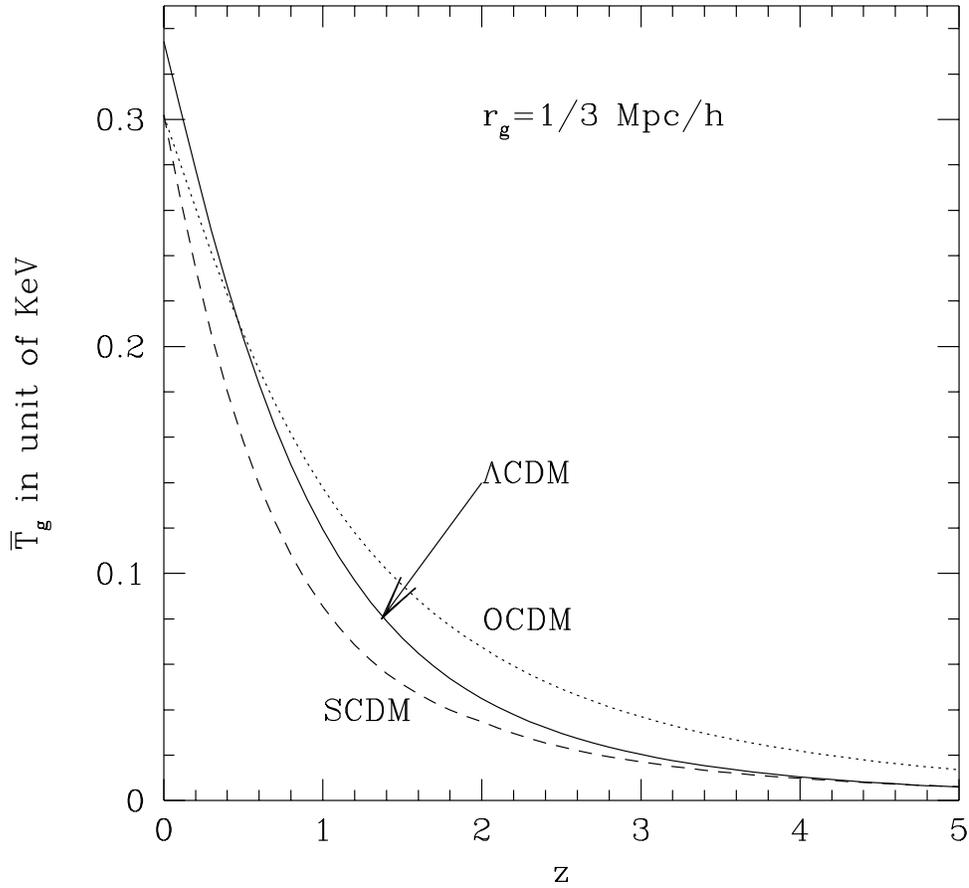}
\caption{The density averaged temperature defined as $\bar{T}_g\equiv
\langle \rho T \rangle/\bar{\rho}$. We assume that the gas overdensity
is the convolution of the dark matter overdensity with a top hat window of
radius $r_g=1/3$ Mpc/h. The gas window function reflects the fact that, due to
the gas pressure, the gas is more diffuse than the dark matter.  }
\end{figure}

\begin{figure}
\label{fig:pressure}
\plotone{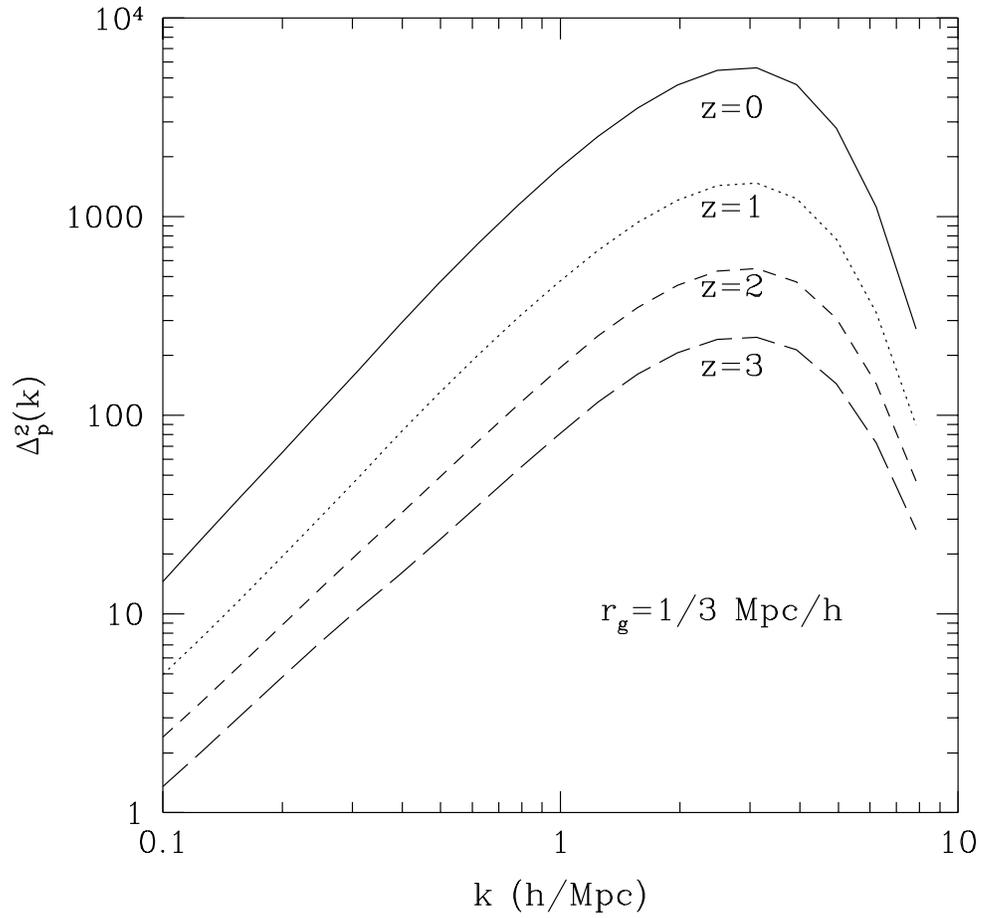}
\caption{The variance of the pressure fluctuation
$\frac{P-\bar{P}}{\bar{P}}$. The peak of the variance is mainly
determined by $r_g$. } 
\end{figure}

\begin{figure}
\label{fig:cl}
\plotone{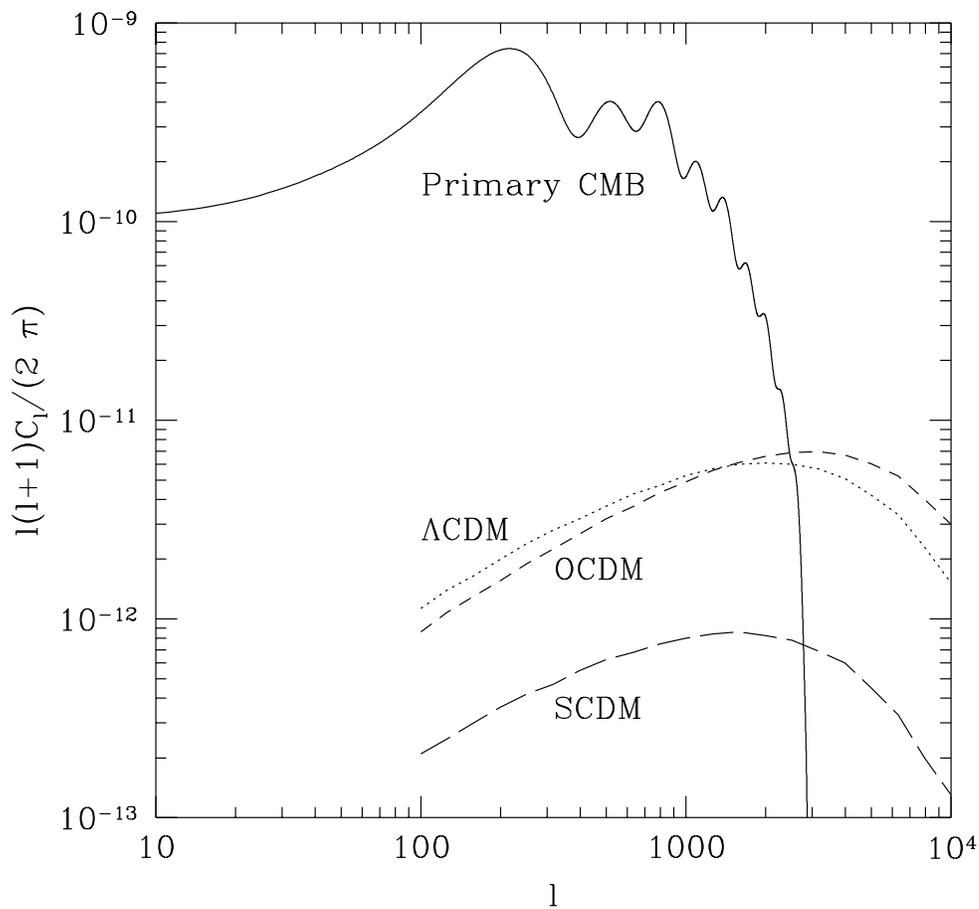}
\caption{The SZ angular power spectrum. The SCDM has a much smaller
$\sigma_8 \sim 0.5$. Due to the strong dependence of $C_l$ on
$\sigma_8$, the angular power spectrum of a SCDM cosmology is much
smaller than the Open CDM model and the $\Lambda$CDM model. }
\end{figure}

\begin{figure}
\label{fig:szcon}
\plotone{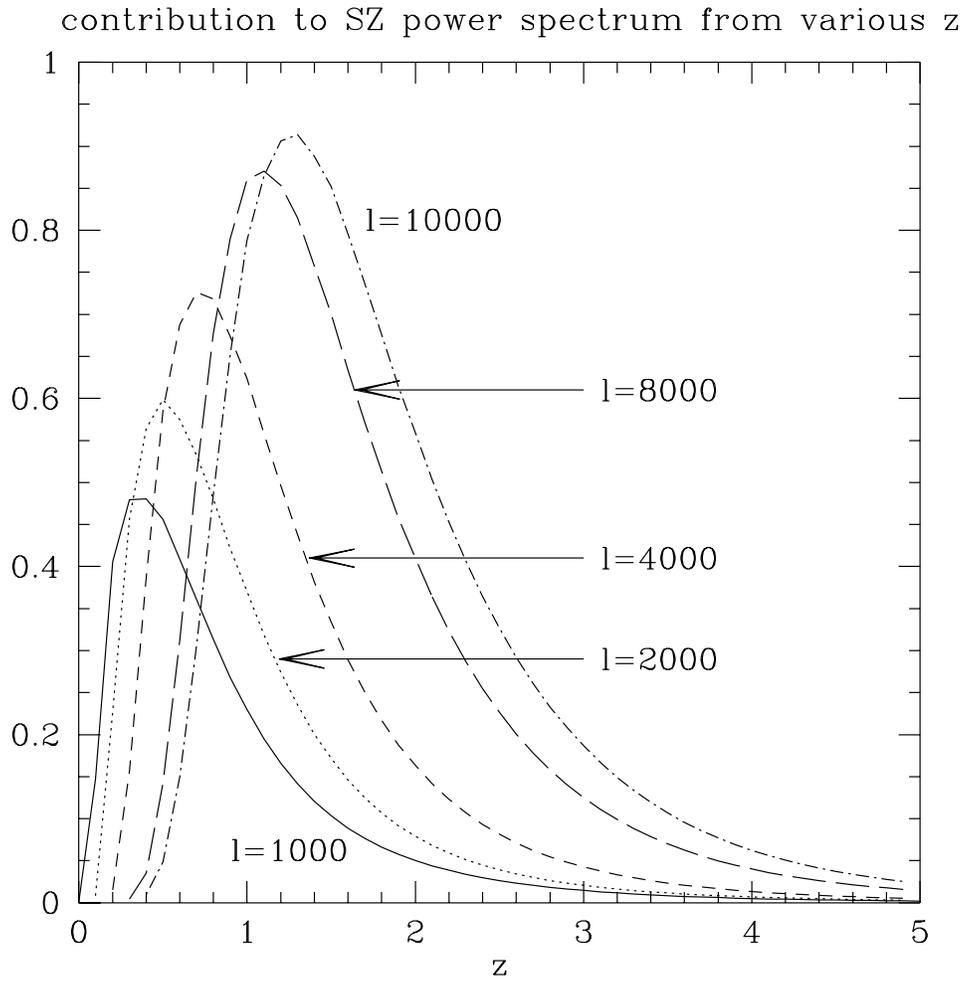}
\caption{Redshift dependence of $C_l$. The typical peak contribution
of IGM to SZ effect is from $z\sim 1$. Smaller peak probes more
distant universe.}
\end{figure}

\begin{figure}
\label{fig:heating}
\plotone{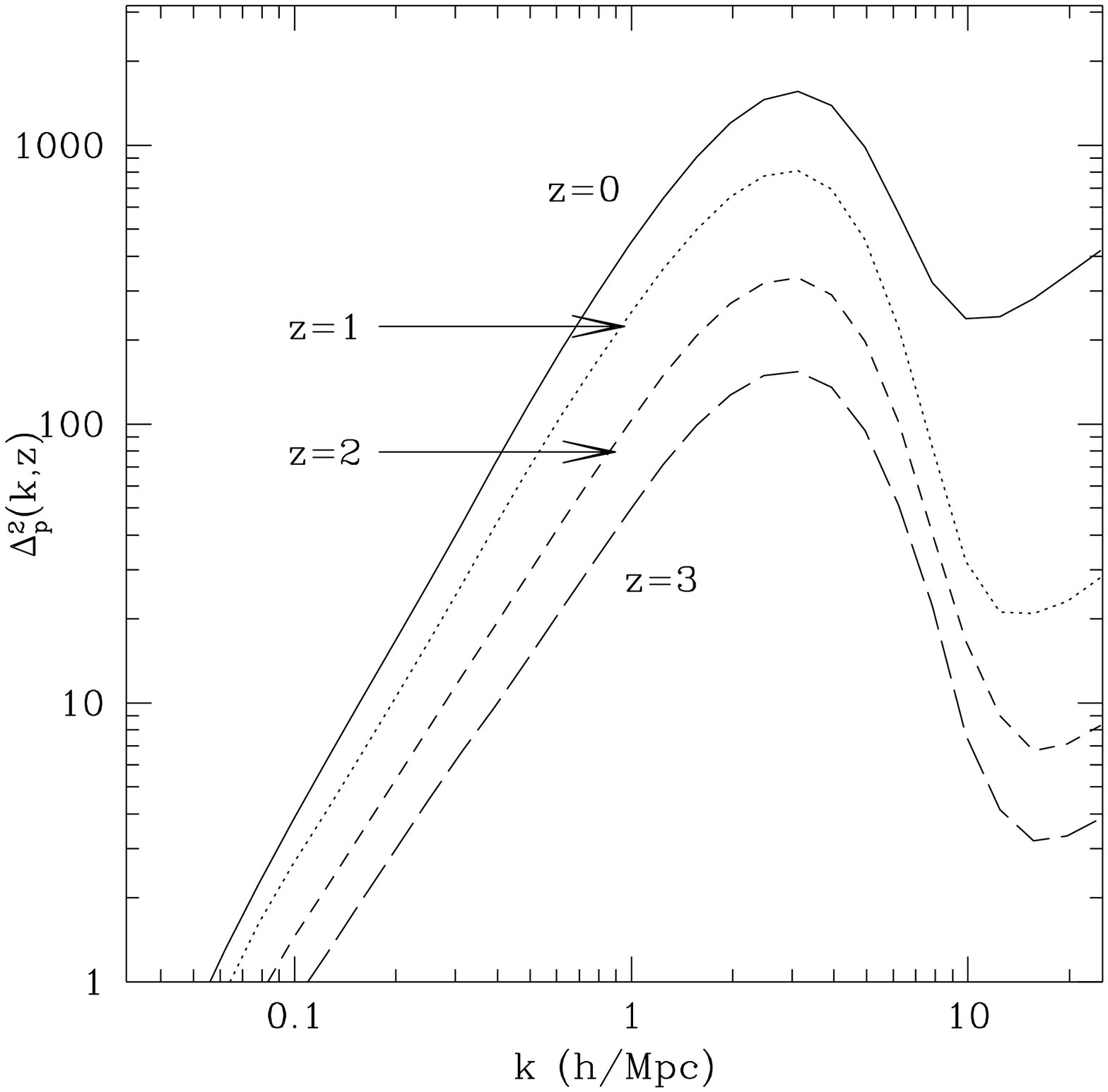}
\caption{Effect of non-gravitational heating if not correlated with
the gravitational heating.  The uprising tail at large $k$ (smaller
distance) is the distinct signature of this non-gravitational
heating. The power spectrum contributed by the non-gravitational
heating is a constant. So, $\Delta^2_p(k)\propto k^3$ and becomes
dominant on scales smaller than that of the gravitational heating peak.}
\end{figure}


\begin{references}
\reference Persic, M. \& Salucci, P., 1992, \mnras, 258, 14P
\reference Fukugita, M., Hogan, C.J. \& Peebles, P.J.E., 1998, \apj, 503, 518
\reference Array for Microwave Background Anisotropy 2003,
           http://www.asia a.sinica.edu.tw/amiba/  
\reference Perna, R. \& Loeb, A., 1998, \apjl, 503, L137
\reference Zhang,P.J. \& Pen, U.L., 2001, \apj, 549, 18
\reference Zel'dovich, Y.B. and Sunyaev, R., 1969, Ap\&SS, 4, 301
\reference Cooray, A., Hu, W.  \& Tegmark, M., 2000, \apj, 540, 1
\reference Sloan Digital Sky Survey, http://www.sdss.org
\reference Komatsu, E. \& Kitayama, T., 1999, \apjl, 526, L1
\reference Fry, J.N., 1984, \apj, 279, 499
\reference Scoccimarro, R. and Frieman, J., 1999, \apj, 520, 35
\reference Pen, U.L., 1999, \apjl, 510, L1
\end{references}
\end{document}